# Coexistence via trophic cascade in plant-herbivore-carnivore systems under intense predation pressure

Mozzamil Mohammed[*,1], Mohammed AY Mohammed[2], Abdallah Alsammani[3], Mohamed Bakheet[4], Cang Hui[5,6,7], Pietro Landi[5,7]

1. Odum School of Ecology, University of Georgia, Athens, GA, USA
2. Department of Mathematics and Statistics, Georgia State University, Atlanta, GA, USA
3. Department of Mathematics, Jacksonville University, Jacksonville, FL, USA
4. Department of Infectious Diseases, University of Georgia, Athens, GA, USA
5. Department of Mathematical Sciences, Stellenbosch University, Stellenbosch 7602, South Africa
6. African Institute for Mathematical Sciences, Cape Town 7945, South Africa
7. National Institute for Theoretical and Computational Sciences (NITheCS), South Africa

## Abstract

Carnivores interact with herbivores to indirectly impact plant populations, creating trophic cascades within plant-herbivore-carnivore systems. We developed and analyzed a food chain model to gain a mechanistic understanding of the critical roles carnivores play in ecosystems where plants face intense herbivory. Our model incorporates key factors such as seed production rates, seed germination probabilities, local plant interactions, herbivory rates, and carnivore predation rates. In the absence of carnivores, herbivores significantly reduce plant densities, often driving plants to extinction under high herbivory rates. However, the presence of carnivores suppresses herbivore populations, allowing plants to recover from herbivore pressure. We found that plant densities increase with carnivore predation rates, highlighting top-down effects and underscoring the importance of conserving carnivores in ecosystems where plants are at high risk of extinction from herbivory. Our results also show that carnivore density increases with seed-production rates, while herbivore density remains constant, indicating that plants benefit carnivores more than herbivores. This increase in carnivore density driven by high seed-production rates reflects bottom-up effects in the system. Overall, our study demonstrates that plants, herbivores, and carnivores can coexist even under intense predation stress. It suggests that carnivores play a crucial role in regulating plant and herbivore populations, with significant potential for maintaining biodiversity within ecosystems.

[*] Correspondence: mozzamil.mohammed@uga.edu

# Introduction

Natural plant communities are often shaped by the presence of various animal species, particularly those that consistently feed on plants (Schmitz et al., 2000; Burkepile and Parker, 2017). Interactions between plants and their natural enemies are crucial in determining plant abundance and can significantly impact overall biodiversity (Koerner et al., 2018). Herbivores, as key natural enemies, influence plant populations' short-term and long-term dynamics (Feng and DeAngelis, 2017; Mohammed, 2024). These herbivore-plant interactions can significantly reduce plant abundance while enabling herbivore persistence. However, the impact of herbivores on plants can be mitigated by the plants' physical and chemical defenses (Feng et al., 2008; Zhong et al., 2021) and by the presence of secondary predators that feed on herbivores (Strong, 1992; Schmitz et al., 2000). For example, carnivores can indirectly reduce plant mortality caused by herbivores by diminishing the predation pressure that herbivores exert on plants.

The indirect effects of carnivores on plant populations have been widely debated (Schmitz et al., 2000; Strong, 1992). It is well recognized that the introduction or removal of carnivores can trigger trophic cascades in plant-herbivore-carnivore systems. A trophic cascade occurs when carnivores reduce the growth of their herbivore prey, thereby lessening the predation pressure that herbivores exert on plants (Health et al., 2014; Ripple et al., 2016). Carnivores directly influence herbivore abundance through predator-prey interactions (Sih et al., 1985), which can benefit plant populations by mitigating the negative impact of herbivores on plants. Traditionally, trophic cascades are considered less common in terrestrial ecosystems compared to aquatic systems (Strong, 1992). However, plants also develop their own defense mechanisms to reduce herbivore predation, in addition to the effects of carnivores on herbivores. Thus, the potential influence of herbivores on plants is shaped by both plant defense strategies and the presence of carnivores. This introduces additional complexity in understanding the indirect impact of carnivores on plants. One way to manage this complexity is to separately examine the effects of plant defense strategies and carnivores on plant-herbivore interactions, and then compare their relative impacts. This approach allows for a more accurate assessment of how each factor affects plant populations. While substantial research has focused on plant defense strategies and their influence on plant-herbivore interactions (Li et al., 2006; Feng et al., 2008; Burkepile and Parker, 2017; Zhi-Wei et al., 2021), this paper primarily aims to assess the indirect impact of carnivores on plant populations and explore how trophic cascades are triggered within plant-herbivore-carnivore systems.

We develop and analyze a food chain model to explore the direct interactions between plants and herbivores, and between herbivores and carnivores through predator-prey dynamics. Our goal is to elucidate the indirect effects of carnivores on plant populations that experience herbivory, and to identify the conditions that enable the coexistence of plants, herbivores, and carnivores. This novel model, developed using the pair-approximation method (Harada and Iwasa, 1994; Dieckmann et al., 2000; Payne, 2019; Mohammed et al., 2023), provides a more accurate description of plant dynamics compared to classic mean-field approximation models. Our model incorporates crucial aspects of plant dynamics, including seed production rates, seed germination probabilities, and local interactions among plant individuals. It effectively captures both the local and global dynamics of plants and the global dynamics of herbivores and carnivores. The predation rates of herbivores and carnivores are integral components of our model, directly or indirectly influencing plant populations. By varying these predation rates, we investigate the ecological impact of carnivores on plants across a range of herbivory rates, from low to high. Our findings highlight the critical role of carnivores in facilitating species coexistence in plant-herbivore-carnivore systems under intense herbivory pressure.

## Method

### Pair-approximation model of plant-herbivore-carnivore interactions

Pair approximation is a modelling approach that explicitly accounts for local interactions among distributed individuals (Dieckmann et al., 2000; Harada and Iwasa, 1994). It has been employed to approximate the spatial dynamics of plants and plant-animal interactions in spatially explicit landscapes (Harada and Iwasa, 1994; Mohammed et al., 2023). The pair approximation considers three probability densities: the local and global densities as well as the densities of adjacent pairs. Using the pair-approximation method, we develop a process-based model of plant-herbivore-carnivore interactions to investigate the role of carnivores for the persistence of plants. In a lattice context, we assume that the landscape consists of grid cells (also called sites), with each cell capable of at most one plant individual.

The global density of plants, denoted as $P_+$, is the proportion of occupied grid cells, $0 \leq P_+ \leq 1$, and represents the probability that a randomly chosen grid cell is occupied by a plant. The probability that a randomly chosen site is empty is thus given by $P_0 = 1 - P_+$. The local density of plants, denoted as $q_{+|+}$, is defined as the conditional probability that a randomly chosen site in the neighborhood of a focal plant is also occupied by a plant. It can be computed as $q_{+|+} = P_{++}/P_+$, where $P_{++}$ is the joint probability of a randomly chosen pair of neighboring sites both being occupied by plants. The conditional probability that a randomly chosen neighboring site of an occupied site is empty is, therefore, $q_{0|+} = 1 - q_{+|+}$. We assume that plant germination in an empty site requires that it is adjacent to at least a plant (i.e., to allow local seed dispersal). Comparing local density against global density describes the spatial clustering of plants (Hui et al., 2006; Hui, 2009): i.e., plants are segregated if the local density is less than the global density ($q_{+|+} < P_+$), aggregated if the local density is greater than the global density ($q_{+|+} > P_+$), and randomly distributed if the local and global densities are equal ($q_{+|+} = P_+$).

Plant distributions are typically influenced by herbivores in their environment, such as seed predators. We consider herbivores to be granivores, causing seed mortality through herbivory. In the presence of herbivores, only a fraction of seeds can survive. The fate of a dispersed seed is thus, either eaten by herbivores or dispersed into a neighbouring site around the site of its parental plant. Let $r$ be the production rate of dispersible seeds per plant (a compound rate of seed production and dispersal), $c$ the rate of seeds being consumed by herbivores and $s$ the rate of seeds being dispersed into neighboring sites. For simplicity, we ignore the seeds that landed in the site where they were produced, and therefore, the fraction of seeds predated by herbivores is $\frac{c}{c+s}$, and the fraction of seeds dispersed is $\frac{s}{c+s}$. The germination rate of seeds surviving predation is determined by the product of the probability that these dispersed seeds find an empty site in the neighborhood $q_{0|+}$ and the intrinsic germination probability of seeds $g$. These probabilities give the recruitment rate of plants, $g q_{0|+} \frac{s}{c+s} r P_+$. In addition to natural mortality ($d_P$), seed predation can result in additional plant mortality at rate $\kappa \frac{c}{c+s} r P_+$, where $\kappa$ is the intensity of byproduct plant mortality. To this end, the dynamics of the global plant density are described by following differential equation

$$\dot{P}_+ = g q_{0|+} \frac{s}{c+s} r P_+ - (d_P + \kappa \frac{c}{c+s} r) P_+. \tag{1}$$

Without predators ($c = 0$), Eqn. (1) can be written in the form of the logistic equation when $q_{0|+} = 1 - P_+$, corresponding to the mean-field approximation model where plants are randomly distributed in space. Based on the conditional probability definition, we have $\dot{q}_{+|+} = -P_{++}\dot{P}_+/P_+^2 + \dot{P}_{++}/P_+$, and after some algebra, we obtain final equation that describes the local plant density dynamics (Appendix B; Mohammed et al., 2018, 2023) :

$$\dot{q}_{+|+} = g q_{0|+} \frac{s}{c+s} r P_+ \left( \frac{2-n q_{+|+}}{n P_+} + \frac{2(n-1)}{n} \frac{q_{0|+}}{P_0} \right) - \left( d_P + \kappa \frac{c}{c+s} r \right) q_{+|+}, \tag{2}$$

where $n$ is a positive integer, representing the number of neighbouring sites of a focal plant.

We assume herbivores are specialists and cannot survive in sites that are absent of plants. The dynamics of herbivores are driven by seed predation and the herbivore mortality rate $d_H$. In the absence of plants, the herbivores population decays exponentially, and they eventually go extinct. The rate of seed predation is formulated as, $c = a_H H$, where $a_H$ is the per-capita herbivory rate and $H$ is the herbivore density (assuming herbivores are roaming across the entire landscape). Herbivores dynamics are also influenced by the presence of carnivores which suppress herbivores growth by feeding on them. Carnivores are important examples of secondary consumers that affect plant-herbivore interactions. We consider herbivore mortality by carnivores through predation, described using the Holling Type II functional response ($\frac{a_Z H}{1+a_Z \tau_Z H}$, where $a_Z$ is the carnivore per-capita attack rate and $\tau_Z$ is the handling time). Note that, we assume that carnivores need much more time to handle herbivores than the negligible time required by herbivores to handle plant seeds. Thus, we neglect the herbivore handling time in our model. We denote by $Z$ the density of carnivores. The herbivore density dynamics are described by

$$\dot{H} = \beta_H \frac{c}{c+s} r P_+ - d_H H - \frac{a_Z H}{1+a_Z \tau_Z H} Z, \qquad (3)$$

where $\beta_H$ is the herbivore conversion efficiency, $0 < \beta_H < 1$. The first term represents the positive contribution of primary producers (plants) to the overall growth of herbivore predators. The second term corresponds to the natural death of herbivores. The last term corresponds to the herbivore mortality caused by predation by carnivores.

Carnivores are assumed to be specialists, i.e., they entirely rely on herbivores for food and become extinct in their absence. The dynamics of carnivores are influenced by their natural mortality rate $d_Z$ and the carnivore predation rate and the efficiency of carnivores to covert the predated herbivores into their biomass ($0 < \beta_Z < 1$). In the absence of herbivores, carnivores go extinct exponentially. The carnivore density dynamics are described by

$$\dot{Z} = \beta_Z \frac{a_Z H}{1+a_Z \tau_Z H} Z - d_Z Z. \qquad (4)$$

The first term represents herbivore predation's positive contribution to the carnivore predators' overall growth. The second term corresponds to the natural death of carnivores.

The system described by Eqns. (1) - (4) involves a description for the local and global dynamics of plants and the population dynamics of herbivores and carnivores, and they yield a full dynamical system for plant-herbivore-carnivore interactions. The numerical analysis of the model is implemented in R using the ode45 solver of the R deSovle package. Parameter values used in the simulation are provided in Table 1.

**Table 1:** Variable and parameter descriptions and parameter values used in the simulation.

| Symbol | Description | Range/value |
|---|---|---|
| $P_+$ | Global density of plants | [0,1] |
| $q_{+\|+}$ | Local density of plants | [0,1] |
| $H$ | Herbivore density | $[0, \infty)$ |
| $Z$ | Carnivore density | $[0, \infty)$ |
| $r$ | Seed-reproduction rate | [1,40] |
| $d_P$ | Plant natural death rate | 0.6 |
| $s$ | Seed dispersal rate | [1,40] |

| $g$ | Seed germination probability | 0.3 |
| $d_H$ | Herbivore natural death rate | 0.1 |
| $a_H$ | Herbivore per-capita attack rate | [0,40] |
| $\kappa$ | Intensity of byproduct plant mortality | [0.1,1] |
| $\beta_H$ | Herbivore conversion efficiency | 0.5 |
| $d_Z$ | Carnivore natural death rate | 0.1 |
| $a_Z$ | Carnivore per-capita attack rate | [0,40] |
| $\beta_Z$ | Carnivore conversion efficiency | 0.5 |
| $\tau_Z$ | Handling time for carnivores | 0.1 |
| $n$ | Number of neighbouring sites | 4 |

# Results

## Herbivores typically decrease plant densities and create transient behaviour

In the absence of carnivores, herbivory pressure initially pushes plants towards low density levels, resulting in both plants and herbivores remaining at low densities for a considerable amount of time (Fig. 1). This low density of herbivores enables plants to increase their abundance, reaching a higher-density transient state (Fig. 1b, time between approximately 750 to 800). An oscillating second transient phase for both plants and herbivores follows (Fig. 1b, time between approximately 800 to 1100), before approaching the long-term attractor. The reason for the first transient phase is the reduced herbivory pressure on plants, allowing plant density to increase. This triggers the second phase, where herbivores respond to the increase in plant density by increasing their own density (Fig. 1b and c). As herbivores increase in density, herbivory pressure on plants resumes, pushing plant density downward and causing fluctuations in the densities of both plants and herbivores. Eventually, plants and herbivores could coexist at equilibrium for a given level of herbivory. The influences of herbivores on the transient and long-term dynamics of plants are studied in detail in Mohammed et al. (2024). Our results demonstrate that local plant density is higher than global density, indicating the spatial aggregation of plants. This is driven by the model assumption that plants disperse their seeds locally within their own neighbourhoods.

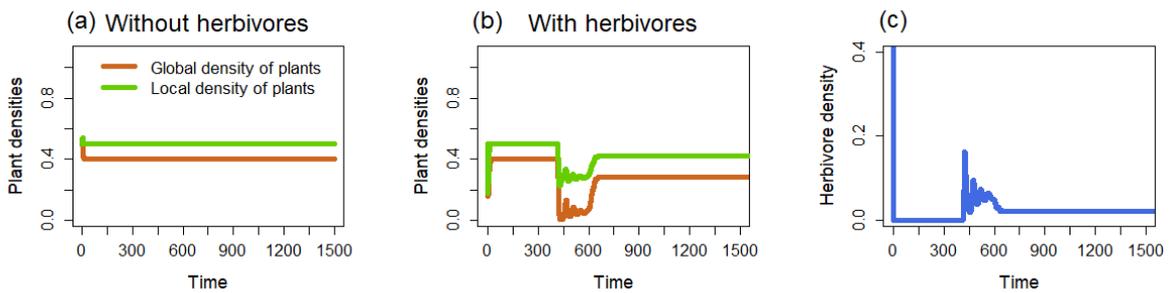

**Fig. 1:** Herbivores typically decrease the plant densities in the absence of carnivores. (a) Changes in the local and global densities of plants over time, with the x-axis representing the simulation time and the y-axis representing plant densities in the absence of herbivores. (b) Changes in the local and global densities of plants over time, with the x-axis representing the simulation time and the y-axis representing plant densities in the presence of herbivores. (c) Changes in the density of herbivores over time, with the x-axis representing the simulation time and the y-axis representing herbivore density. Parameters used are: $d_P = 0.6$, $d_H = 0.1$, $r = 4$, $s = 10$, $g = 0.3$, $a_H = 10$, $\alpha_H = 0.5$, $z = 4$.

**Carnivores increase plant density via top-down effects**

The introduction of carnivore species into the plant-herbivore system significantly influenced plant-herbivore interactions. Carnivores triggered top-down effects, allowing plants to remain at high densities despite herbivory pressure (Fig. 2). We found that a high density of carnivores suppressed herbivore growth, enabling plants to increase their density (Fig. 2a and b). Although plants were abundant for a long period, both herbivores and carnivores were at low densities during their initial transient phase (Fig. 2b, time between approximately 100 to 400). Their dynamics were slower compared to plant dynamics. Herbivores increased from low to high density, followed by an increase in carnivore density in response to the herbivore density rise. While plants suffered from herbivory pressure, the herbivore density declined due to the low availability of food and the pressure exerted by carnivores. Importantly, in the absence of carnivores, plants and their herbivore predators could coexist, but with carnivores, plants maintained higher densities at equilibrium (compare final attractor in Fig. 1b and 2a). These findings conclude that increases in plant density are enabled by carnivore predators via top-down effects in this particular system.

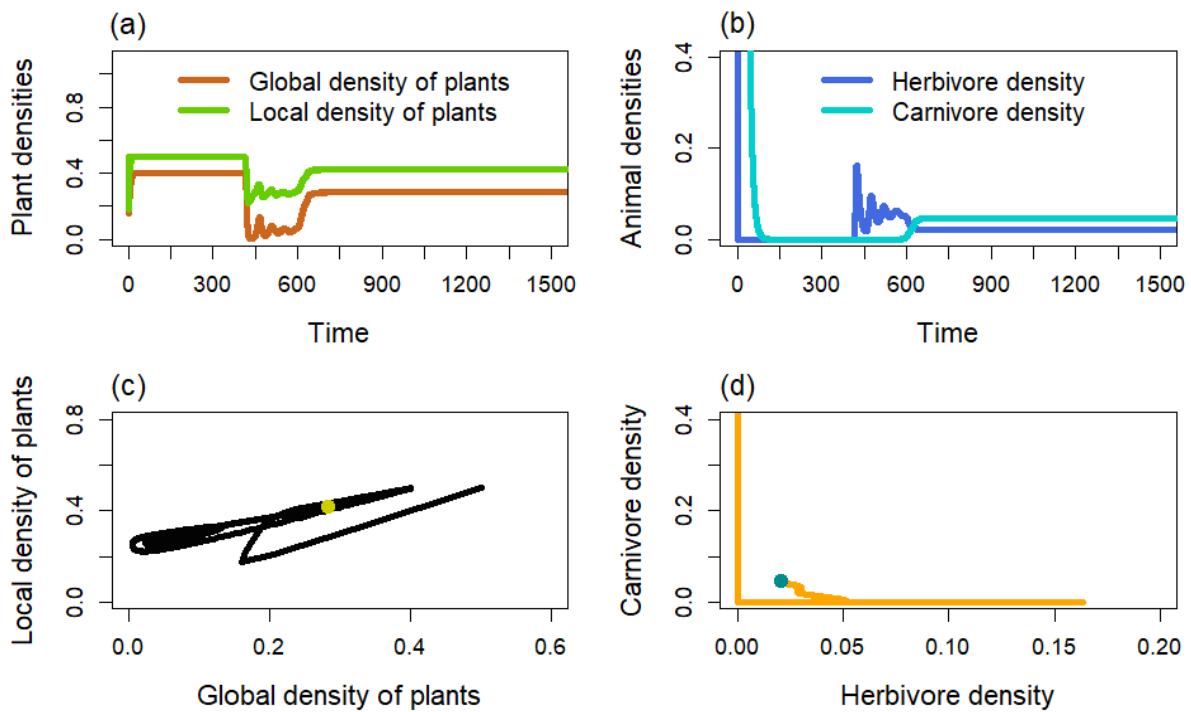

**Fig. 2:** Carnivores increase plant density via top-down effects. The dynamics of local and global plant densities (a), herbivore dynamics (b), and the corresponding plant (c) and animal (d) trajectories on the phase plane are shown. In panel (a), the x-axis represents simulation time, and the y-axis represents plant densities. In panel (b), the x-axis represents the simulation time, and the y-axis represents herbivore density. In panel (c), the x-axis represents the global density of plants, and the y-axis represents the local density of plants. In panel (d), the x-axis represents herbivore density, and the y-axis represents carnivore density. The filled circles in panels (c) and (d) indicate equilibrium points. Parameters used are: $d_P = 0.6$, $d_H = 0.1$, $d_Z = 0.1$, $r = 4$, $s = 10$, $g = 0.3$, $a_H = 10$, $a_Z = 10$, $\alpha_H = 10$, $\alpha_Z = 10$, $\tau_Z = 0.1$, and $z = 4$.

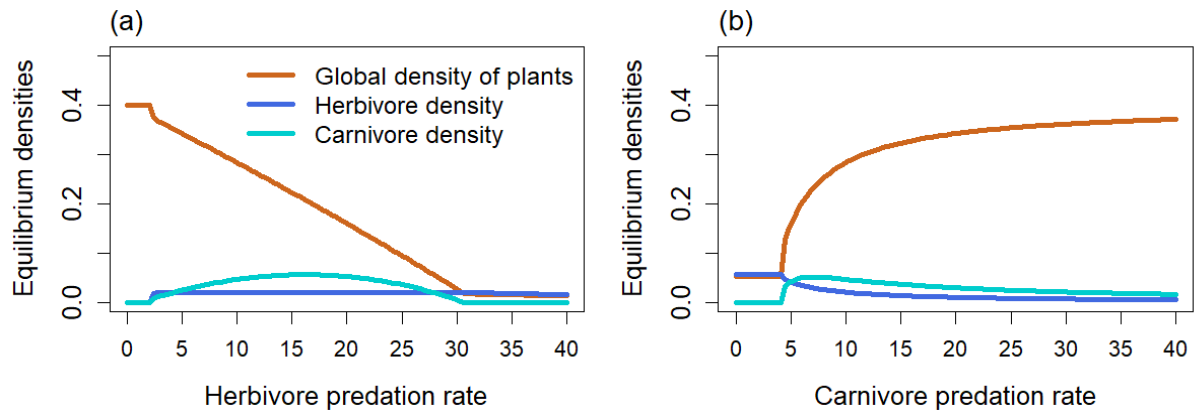

**Fig. 3:** Increases in the carnivore predation rate enable plants to increase in density. Changes in the equilibrium densities as functions of the herbivore predation rate (a) and the carnivore predation rate (b) are shown. In panels (a) and (b), the x-axis represents the predation rates of herbivores and carnivores, respectively, while the y-axis represents the equilibrium densities of plants, herbivores, and carnivores. Parameter values are the same as in the caption of Figure 2.

### Increases in the carnivore predation rate enable plants to increase in density

We systematically varied the herbivore and carnivore predation rates, both independently and in combination, to elucidate their influences on the equilibrium densities of plants. We found that both herbivores and carnivores go extinct when their predation rates are low (Fig. 3). In our model, carnivores are assumed to completely rely on herbivores for food, so they go extinct whenever herbivores go extinct. Similarly, herbivores go extinct whenever plants go extinct. For carnivores to persist, herbivores must have a sufficient herbivory rate. Herbivores are suppressed by carnivores even when the herbivore predation rate is high and carnivores are at low density. At exceedingly high herbivory rates, both plants and herbivores persisted at low densities, leading to the extinction of carnivores. Interestingly, carnivores go extinct when plants reach a specific abundance level, even if herbivores are still present in the system. One possible reason is that the percentage of mass and energy passed to carnivores from plants through herbivores is insufficient to sustain carnivores.

### Coexistence under intense predation pressure

Plant density increases with the carnivore predation rate, while herbivore density decreases with increased predation by carnivores (Fig. 4a and b). The decrease in herbivore density intuitively leads to a decrease in carnivore density, even at high carnivore predation rates (Fig. 4b and c). The interplay between herbivore and carnivore predation rates demonstrated a coexistence scenario of the three species under high predation pressure. This scenario is only possible when the degree of herbivory is matched by the degree of carnivore predation. In other words, if herbivores are aggressively attacking plants, carnivores must also be aggressive towards herbivores. Carnivores with low predation rates cannot coexist with plants and herbivores when herbivory pressure is high (Fig. 4c). Strong herbivory pressure will reduce plant density to low levels, maintaining only plants and their herbivore predators at low densities. These low density levels are insufficient to support the persistence of carnivores, leading to their exclusion from the system.

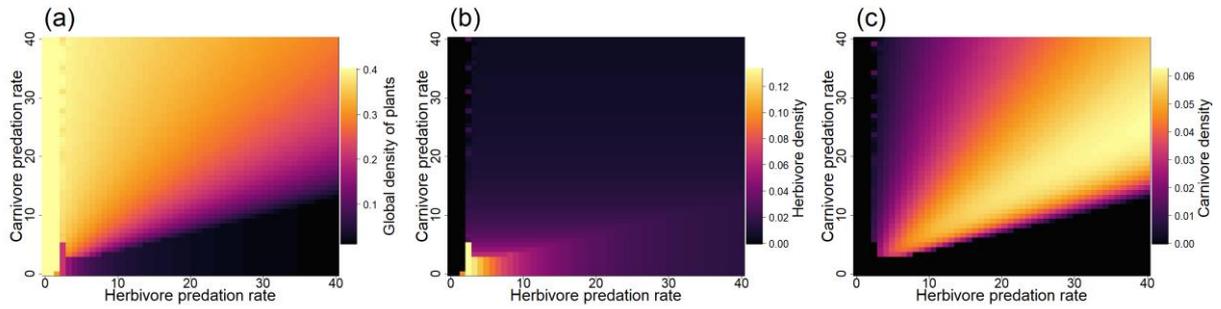

**Fig. 4:** Coexistence under intense predation pressure in a plant-herbivore-carnivore system. Changes in the equilibrium densities of plants (a), herbivores (b), and carnivores (c) are shown, color-coded as functions of herbivore predation rates (x-axis) and carnivore predation rates (y-axis). Parameter values are the same as in the caption of Figure 2.

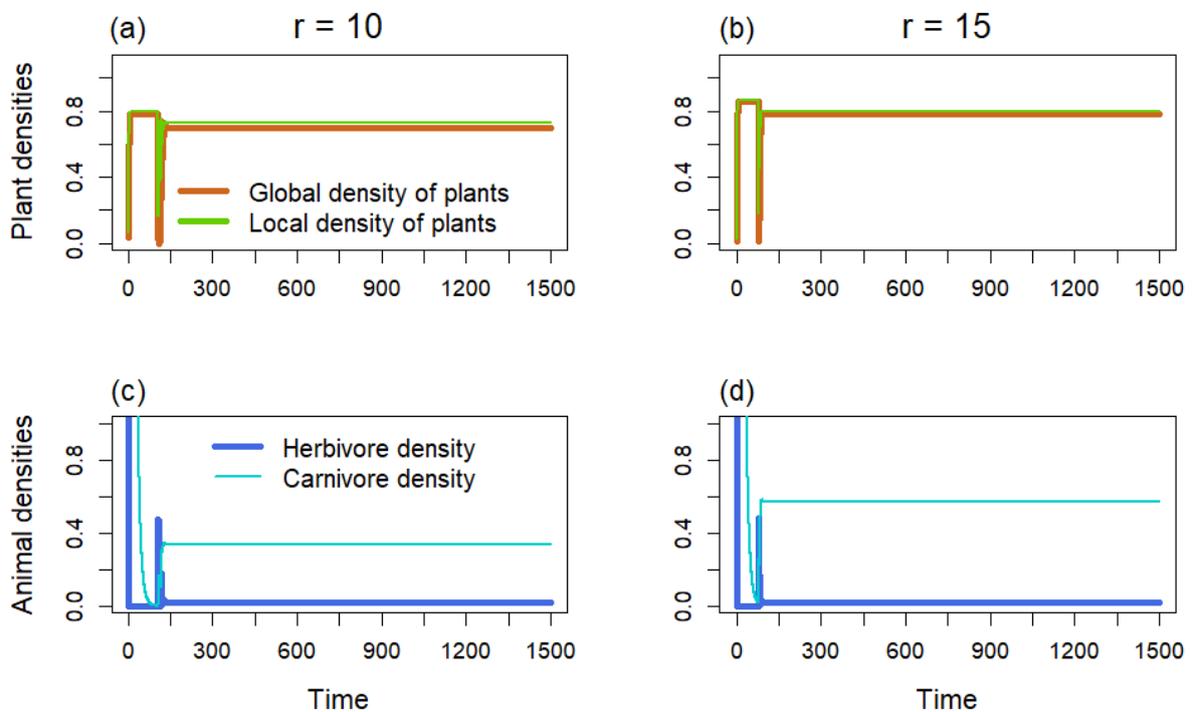

**Fig. 5:** Carnivore density increases with seed-production rate via bottom-up effects. The dynamics of local and global density of plants (a) and (b), and densities of herbivores and carnivores (c) and (d). Dynamics are shown for different values of seed-production rates, $r = 10$ (the first column, panels a,c) and $r = 15$ (the second column, panels b,d). In all panels, the x-axis represents time, and the y-axis represents plant densities (a and b) and animal densities (c and d). Parameter values are the same as in the caption of Figure 2.

**Carnivore density increases with seed-production rate via bottom-up effects**

We further varied the seed-production rate and found that it has a strong indirect influence on carnivore density. Specifically, carnivore density increases with the seed-production rate via bottom-up effects (Figs. 5 and 6a). High seed-production rates shorten the transient phase of plants and herbivores, allowing herbivores to quickly increase from low abundance (Fig. 3; see Mohammed et al. 2024). The rapid increase in herbivore density pushed plants to low density but enabled carnivore density to rise. The increase in carnivore density then pushed herbivore density downward, allowing plants to recover. Interestingly, while carnivores caused herbivore density to drop to very low levels, both plants and carnivores remained at high densities.

Although herbivores are the only food source for carnivores, high seed-production rates allowed carnivores to maintain high densities after suppressing herbivores. We also varied the natural dispersal rate of plants to investigate its influence on herbivores and carnivores. Increases in plant dispersal ability typically prevent seed predation by herbivores, potentially decreasing the densities of both herbivores and carnivores (Fig. 6b).

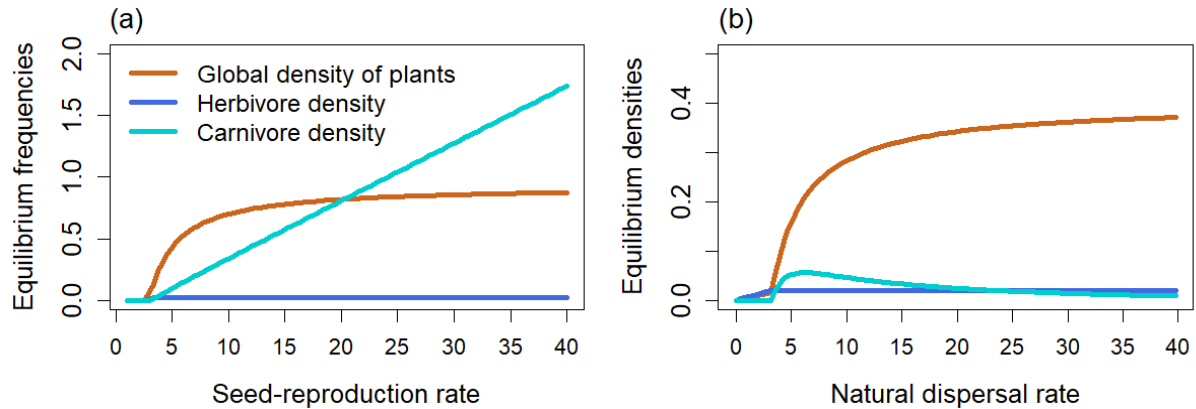

**Fig. 6:** Carnivore density increases with seed-production rate via bottom-up effects. Changes in the equilibrium densities as functions of the seed-reproduction rate (a) and the natural dispersal rate of plants (b) are shown. In panels (a) and (b), the x-axis represents the seed-reproduction rate and the natural dispersal rate, respectively, while the y-axis represents the equilibrium densities of plants, herbivores, and carnivores. Parameter values are the same as in the caption of Figure 2.

Finally, we assessed the impact of byproduct plant mortality due to herbivory on the system dynamics. The results demonstrated that changes in byproduct plant mortality do not influence the equilibrium densities of plants and consequently the densities of herbivores and carnivores (Appendix A). From a modelling perspective, incorporating byproduct plant mortality adds more realism to the model, but it does not affect seed removal by herbivores.

## Discussion

Plants and herbivores are key species that regulate the flow of mass and energy across trophic levels within ecosystems. Herbivores directly affect plants by feeding on them, while carnivores indirectly influence plants via trophic cascades (Silliman and Angelini, 2012; Ripple et al., 2016). Predator-prey dynamics between herbivores and carnivores are well understood, but the indirect effects of carnivores on plants have been debated, resulting in two ecological paradigms (Strong, 1992; Schmitz et al., 2000; Peterson et al., 2014). The first paradigm suggests that plants defend themselves against herbivores through physical and chemical mechanisms, implying minimal ecological impact on plants from carnivores. The second paradigm posits that herbivores significantly impact plants and that carnivores must mitigate these negative effects to maintain ecosystem functioning. However, the indirect effects of carnivores on plants should be system-specific, making a universal rule for such effects unfeasible (Peterson et al., 2014). While trophic cascades have been empirically studied, there is a lack of mechanistic understanding of how carnivores could indirectly affect plants (Ford and Goheen, 2015). This work aims to gain a mechanistic understanding of the indirect impact of carnivores on plants and how trophic cascades enable the coexistence of all three species under intense predation pressure. Our study is not limited to a specific plant-herbivore-carnivore system but is a general theoretical investigation that can be applied to various terrestrial ecosystems.

Our findings demonstrate that carnivores are crucial for the long-term survival of plants under herbivory pressure. Without carnivores, aggressive herbivores could cause plant extinctions or significantly reduce plant population densities (Zhong et al., 2021; Mohammed, 2024). Herbivores consuming seeds effectively remove potential new plant recruits and can damage plants in the process, leading to additional plant mortality. To prevent herbivores from negatively impacting plants, herbivore growth must be suppressed by either plant toxicity, carnivores, or both (Schmitz et al., 2000; Feng et al., 2008). An important ecological question is how plants, herbivores, and carnivores coexist in a specific area, especially under intense predation pressure. Our study demonstrated that the predation pressure herbivores exert on plants must be balanced by equivalent predation pressure from carnivores on herbivores to ensure the coexistence of all three species. For instance, if herbivore predation is high and carnivore predation is low, all species will go extinct, particularly when herbivores depend entirely on plants for food and carnivores on herbivores.

Interestingly, despite not directly interacting with plants, carnivores can experience extinction when plant density is low. This low plant density is sufficient to support herbivores but not carnivores. One likely reason is that the energy transfer from plants to herbivores is insufficient to sustain carnivores due to losses during biomass conversion at each trophic level of the food chain. This observation of carnivore extinction at low plant density can be readily tested empirically. However, our findings suggest a positive correlation between carnivore and plant abundances, highlighting that the density of carnivores can be increased via bottom-up control (Ford and Goheen, 2015). For instance, planting more seeds can help carnivores establish from low density. These observations are confirmed in our model results, which demonstrated a positive linear relationship between carnivore density and the seed-production rate. This indicates that the indirect benefits plants receive from carnivores are influenced by plant productivity. The higher the seed production, the higher the plant density maintained by carnivores. Without carnivores, plant densities are at significant risk of extinction due to herbivore predation pressure.

Seed dispersal by animals has significant impacts on plant dynamics (Mohammed et al., 2018 and 2023), which can, in turn, alter interactions between plants, herbivores, and carnivores. It would be interesting to incorporate seed dispersers into this study and investigate their roles in reducing herbivory pressure by dispersing seeds away from herbivores. Additionally, one could consider a scenario where seeds consumed by herbivores can be dispersed and germinate with certain probabilities. In our current study, seeds eaten by herbivores are considered as seed removal, with only a fraction of the removed seeds contributing to plant mortality. This means that if the predated seeds are left uneaten, only a fraction will germinate successfully. The contribution of plant mortality by herbivores is quantified in our model by a constant byproduct mortality of plants.

In conclusion, we developed and analyzed a food chain model to mechanistically understand the indirect effects of carnivores on plants and the coexistence of plants, herbivores, and carnivores under intense predation pressure. Using the pair-approximation method, our model accurately captures local interactions among plants, providing improved predictions of plant dynamics. We argue that the model's predictions are empirically testable and its parameters are potentially measurable. Our study concludes that carnivores play an indispensable role in the survival of plants at risk of extinction from herbivores, through both top-down and bottom-up effects (Health et al., 2014). This highlights the importance of carnivores in maintaining biodiversity within ecological systems.

**Acknowledgments**

MM would like to thank Ulf Dieckmann and Åke Brännström for their insightful discussion on our previous research on the plant-frugivore-human system.

**Data availability**

This study does not include or use any data.

**Funding declaration**

None.

**Conflict of interest**

The authors declare that they have no conflict of interest.

# Appendix

*Appendix A: Supplementary figures*

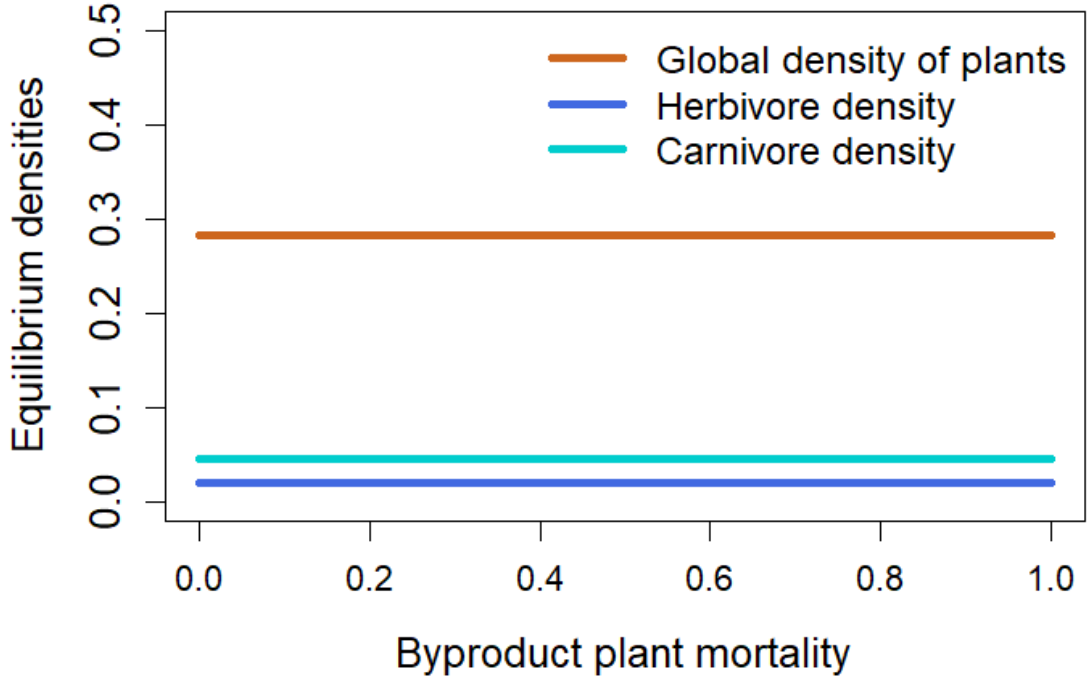

**Fig. A1**: Changes in the equilibrium densities of plants, herbivores, and carnivores as function of the byproduct plant mortality. Parameter values are provided in the caption of Figure 2 (main text).

*Appendix B: Derivation of the joint probability $P_{++}$ and the local plant density equation $q_{+|+}$*

Here, we derive an equation for the local plant density $q_{+|+}$. Given the definition of local plant density $q_{+|+} = \frac{P_{++}}{P_+}$ (see the main text), then the time derivative is given as

$$\dot{q}_{+|+} = -\frac{P_{++}}{P_+^2}\dot{P}_+ + \frac{1}{P_+}\dot{P}_{++}, \tag{B1}$$

where $P_{++}$ is the probability that two randomly chosen neighboring sites are both occupied. The rate of change in the global plant density $\dot{P}_+$ is described in the main text, and we now need to find an equation for the global plant density of pairs $P_{++}$ (Harada & Iwasa 1994), which is given as

$$\dot{P}_{++} = -2d_P P_{++} + 2\frac{gr}{n}\frac{s}{c+s}P_{+0} + 2\frac{gr}{n}(n-1)\frac{s}{c+s}q_{+|0+}P_{+0} - 2\kappa\frac{cr}{c+s}P_{++}, \tag{B2}$$

where $n$ is the number of the nearest neighboring sites. The first term indicates the transition of a (+,+) pair to a (+,0) pair or (0,+) pair; that is where factor 2 comes from. The second and third terms refer to the birth of the non-dispersed seeds. In the second term, an occupied site contributes by a birth of an individual to its nearest-neighboring empty site with transition from (+,0) pair to (+,+) pair or from (0,+) pair to (+,+) pair. The third term, the presence of an occupied site adjacent to the empty site of given nearest-neighboring sites (+,0) may affect the transition of (+,0) to (+,+), that is, the transition from (+,0,+) to (+,+,+) or (0 → +) could be from any of the neighbors of the 0 site. This is why we multiply by $q_{+|0+}$. The pair approximation method neglects the effects of the neighbor-of-the neighbor, therefore $q_{+|0+} \approx q_{+|0}$. We have

$$P_{+0} = P_{0+} = q_{0|+}P_+$$

$$q_{+|0+} \approx q_{+|0} = \frac{P_{0+}}{P_0} = \frac{q_{0|+}P_+}{P_0}$$

The $\dot{P}_{++}$ equation can be written as

$$\dot{P}_{++} = -2d_P P_{++} + 2\frac{gr}{n}\frac{s}{c+s}q_{0|+}P_+ + 2\frac{gr}{n}(n-1)\frac{s}{c+s}\frac{q_{0|+}{}^2 P_+{}^2}{P_0}$$
$$- 2\kappa\frac{cr}{c+s}P_{++}.$$

Using the two equations $\dot{P}_+$ (see the main text) and $\dot{P}_{++}$ and the definition provided above, we can now derive an equation to describe the local plant density as follows:

$$\dot{q}_{+|+} = -\frac{P_{++}}{P_+{}^2}\dot{P}_+ + \frac{1}{P_+}\dot{P}_{++}.$$

$$-\frac{P_{++}}{P_+{}^2}\dot{P}_+ = \left(d_P + \kappa\frac{cr}{c+s}\right)q_{+|+} - gr\frac{s}{c+s}q_{0|+}\,q_{+|+}$$

$$\frac{1}{P_+}\dot{P}_{++} = -2\left(d_P + \kappa\frac{cr}{c+s}\right)q_{++} + 2\frac{gr}{n}\frac{s}{c+s}q_{0|+} + 2\frac{gr}{n}(n-1)\frac{s}{c+s}\frac{q_{0|+}{}^2 P_+}{P_0}.$$

Adding these equations, we get

$$\dot{q}_{+|+} = gq_{0|+}\frac{s}{c+s}rP_+\left(\frac{2-nq_{+|+}}{nP_+} + \frac{2(n-1)}{n}\frac{q_{0|+}}{P_0}\right) - \left(d_P + \kappa\frac{c}{c+s}r\right)q_{+|+}. \quad (B3)$$